\documentclass[11pt,twoside]{article}
\usepackage{asp2004}
\usepackage{epsfig}
\usepackage{lscape}

\def\ifundefined#1{\expandafter\ifx\csname#1\endcsname\relax}

\newif\ifpdf
\ifx\pdfoutput\undefined
   \pdffalse      
\else
   \pdfoutput=1   
   \pdftrue
\fi

\def\la{\mathrel{\hbox{\rlap{\hbox{\lower4pt\hbox{$\sim$}}}\hbox{$<$}}}}
\def\ga{\mathrel{\hbox{\rlap{\hbox{\lower4pt\hbox{$\sim$}}}\hbox{$>$}}}}

\newcommand{\be}{\begin{eqnarray}}
\newcommand{\ee}{\end{eqnarray}}

\ifundefined{ensuremath}\def\ensuremath#1{\relax\ifmmode{#1}}
\else${#1}$\fi\else\relax\fi
\ifundefined{nuc}\def\nuc#1#2{\relax\ifmmode{}^{#1}{\protect\text{#2}}
\else${}^{#1}$#2\fi}\else\relax\fi

\newcommand{\phx}{\texttt{PHOENIX}}

\begin{document}
\title{Early Spectra of Supernovae}
\author{ E.~Baron,\altaffilmark{1,2,3}
Peter E.~Nugent,\altaffilmark{2} David
Branch,\altaffilmark{1} and  Peter
H.~Hauschildt\altaffilmark{4}
}

\altaffiltext{1}{Department of Physics and Astronomy, University of
Oklahoma, 440 West Brooks, Rm.~131, Norman, OK 73019, USA}

\altaffiltext{2}{Computational Research Division, Lawrence Berkeley
  National Laboratory, MS 50F-1650, 1 Cyclotron Rd, Berkeley, CA
  94720-8139 USA}

\altaffiltext{3}{Laboratoire de Physique Nucl\'eaire et de Haute
Energies, CNRS-IN2P3, University of Paris VII, Paris, France}

\altaffiltext{4}{Hamburger Sternwarte, Gojenbergsweg 112,
21029 Hamburg, Germany}

\begin{abstract}
We briefly describe the current version of the \phx\ code. We then
present some results on the modeling of Type II supernovae and show
that fits to observations can be obtained, when account is taken for
spherically symmetric, line-blanketed, expanding atmospheres.  We
describe the SEAM method of obtaining distances to supernovae and
briefly discuss its future prospects.
\end{abstract}

\section{The \phx\ Code}

\phx\ \citep[see][and references therein]{hbjcam99} is a generalized,
stellar model atmosphere code for treating both static and moving
atmospheres. The goal of \phx\ is to be both as general as possible so
that essentially all astrophysical objects can be modeled with a
single code, and to make as few approximations as possible. 
Approximations are inevitable (particularly in atomic data where
laboratory values for most quantities are unknown); however, the
agreement of synthetic spectra with observations across a broad class
of astrophysical objects is a very good consistency check. 
We have modeled 
Planets/BDs \citep{barmanna102,cond-dusty,lhires,latel}, Cool
Stars \citep{nextgen199,nextgen299}, Hot Stars \citep[$\beta$CMa,
$\epsilon$CMa, Deneb][]{aufdenecm98,aufdencma99,aufden02},
$\alpha$-Lyrae, Novae \citep{phhnovetal97,novand86pap}, and all types of
supernovae   \citep[SNe Ia, Ib/c, IIP,
IIb][]{b93j4,l94d01,b94i2,mitchetal87a02}.

\phx\ solves the radiative transfer problem by using the
short-characteristic method \citep{OAB,ok87} to obtain the formal
solution of the special relativistic, spherically symmetric radiative
transfer equation (SSRTE) along its characteristic rays. The
scattering problem is solved via the use of  a
band-diagonal approximation to the discretized $\Lambda$-operator
\citep*{phhs392,ok87,hsb94} as our choice of the approximate
$\Lambda$-operator.  This method can be implemented very efficiently
to obtain an accurate solution of the SSRTE for continuum and line
transfer problems using only modest amounts of computer resources.

We emphasize that \phx\ solves the radiative transfer problem
including a full treatment of special
relativistic radiative transfer in 
spherical geometry for all lines and continua. In addition we enforce
the generalized condition of 
radiative equilibrium in the Lagrangian frame, including
all velocity terms and deposition of energy from radiative decay or
from external irradiation. 

We also include a full non-LTE treatment of most ions, using model
atoms constructed from the data of \citet{kurucz93,cdrom22,cdrom23}
and/or from the CHIANTI and APED databases. The code uses Fortran-95
data structures to access the different databases and model atoms from
either or both databases can be selected at execution time.

Absorption and emission is treated assuming complete redistribution
and detailed 
depth-dependent profiles for the lines. Fluorescence effects are
included in the NLTE treatment.
The equation of state used includes up to 26
ionization stages of 40 elements as well as up to 600 molecules.

\section{Early Supernova Spectra and Distances}

Early spectra of supernovae are important to model
with synthetic spectra. Due to the geometrical dilution, early spectra
tell us about the composition and velocity structure of the outermost
layers of the supernova ejecta. Therefore, they can provide clues to
the progenitor structure of SNe Ia, the primordial composition of all
types of supernovae and the nearby circumstellar
medium. \citet{sn93w103} showed that detailed models of the early
spectra of SN~1993J could reveal the primordial metalicity as well as
the total reddening to the supernova.

\begin{figure}[!ht]
\begin{center}
\includegraphics[height=4cm,width=6cm,angle=90]{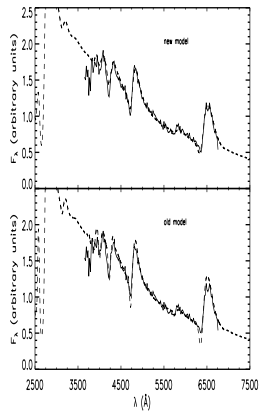}
\caption{A comparison of synthetic spectra of SN~1993W on Aug 20, 1993
  using old (incorrect) and new (correct) NLTE equation of state. The
  change is not dramatic, but H$\alpha$ is better fit with the new
  EOS. \label{fig:93wnewold}}
\end{center}
\end{figure}

With all large computer codes, code improvement is an ongoing
process. We recently improved the NLTE equation of state
and Figure~\ref{fig:93wnewold} compares the results with and without
the improvement. The change is not dramatic and in the right
direction, the newer model fits H$\alpha$ better.

We \citep{bsn99em00} were also able to determine the reddening to
SN~1999em, by modeling the early spectra. We found that $E(B-V) =0.1
\pm 0.05$ for the total (foreground plus parent) color excess.

\begin{figure}[!ht]
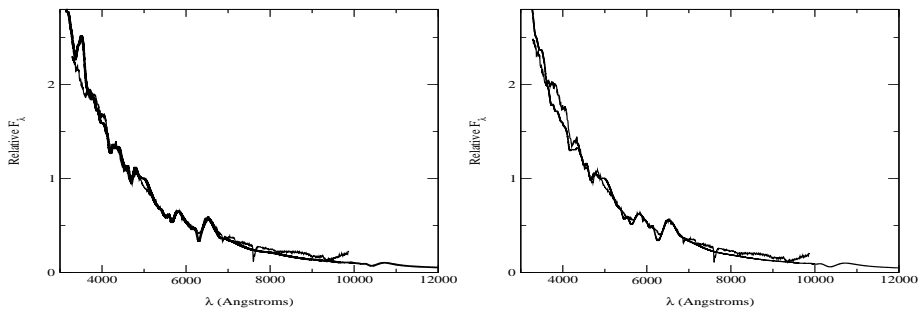

\begin{center}
\includegraphics[height=4cm,width=6cm,angle=0,clip]{baron_fig2a}
\includegraphics[height=4cm,width=6cm,angle=0,clip]{baron_fig2b}
\caption{\label{fig:99em_red} The left panel displays a model with a
  lower temperature and extinction $E(B-V) = 0.05$, while the right
  panel shows a model with a higher temperature and extinction $E(B-V)
  = 0.10$. The Ca II H+K line is very temperature sensitive at this
  early phase and the higher extinction is clearly preferred.}
\end{center}
\end{figure}

In order to determine the values of the fundamental cosmological
parameters, an accurate distance indicator visible to high redshift is
required.  Supernovae are extremely bright and hence can be detected
at cosmological distances with modern large telescopes.  The
reliability of SNe Ia as distance indicators improved significantly
with the realization that the luminosity at peak was correlated with
the width of the light curve \citep{philm15} and hence that SNe~Ia
were correctable candles in much the same way that Cepheids are
\citep{philetal99,goldhetal01,rpk95}. This work and the development of
highly efficient search strategies \citep{perlq097} sparked two groups
to use SNe~Ia to measure the deceleration parameter and to discover
the dark energy \citep{riess_scoop98,perletal99}.

All of the work with SNe~Ia is empirical, based on observed SNe~Ia
template light curves. Another method of determining distances using
supernovae is the ``expanding photosphere method''
 \citep[EPM,][]{kkepm,branepm,eastkir89,esk96} a variation of the
Baade-Wesselink method \citep{baadeepm}. The EPM method assumes
that for SNe~IIP, with intact hydrogen envelopes, the spectrum is
not far from that of a blackbody and hence the luminosity is
approximately given by
$L = 4\pi\,\zeta^2\,R^2\,\sigma\,T^4$,
where $R$ is the radius of the photosphere, $T$ is the effective
temperature,  $\sigma$ is the radiation constant, and $\zeta$ is the
``dilution factor'' which takes into account that in a scattering
dominated atmosphere the blackbody is
diluted \citep*{hlw86a,hlw86b,hw87}. The effective temperature is found
from observed colors, so in fact is a color temperature and not an
effective temperature, the photospheric velocity can be estimated from
observed spectra using the velocities of the weakest lines, 
\(R = v\, t, \)
the dilution factor is estimated from synthetic spectral models,
and $t$ comes from the light curve and demanding self-consistency.

This method suffers from uncertainties in determining the dilution
factors, the difficulty of knowing which lines to use as velocity
indicators, uncertainties between color temperatures and effective
temperatures, and questions of how to match the photospheric radius
used in the models to determine the dilution factor and the radius of
the line forming region \citep{hamuyepm01,leonard99em02}.  In spite of
this the EPM method was successfully applied to SN~1987A in the
LMC \citep{eastkir89,bran87a} which led to hopes that the EPM method
would lead to accurate distances, independent of other astronomical
calibrators. Recently, the EPM method was applied to the very well
observed SN~IIP
1999em \citep{hamuyepm01,leonard99em02,elmhhamdietal99em03}. All three
groups found a distance of 7.5--8.0~Mpc. 
\citet{leonard99em03} subsequently used \emph{HST} to obtain a
Cepheid distance to the parent galaxy of SN~1999em, NGC 1637, and found
$11.7 \pm 1.0$~Mpc.

With \phx, accurate synthetic
spectra of all types of supernovae can be calculated.  The
\textbf{S}pectral-fitting \textbf{E}xpanding \textbf{A}tmosphere
\textbf{M}ethod (SEAM) \citep{b93j3,b94i1,l94d01,mitchetal87a02} was
developed using 
\texttt{PHOENIX}. While SEAM is 
similar to EPM in spirit, it avoids the use of dilution factors
and color temperatures. Velocities are determined accurately by
actually fitting synthetic  and observed spectra. The radius is
still determined by the relationship $R = vt$, (which is an excellent
approximation because all supernovae quickly reach homologous
expansion) and the explosion time is found by demanding self consistency.
SEAM uses all the spectral information available in the observed spectra
simultaneously which broadens the base of parameter determination.
Since the spectral energy distribution is known completely from the
calculated synthetic spectra, one may calculate the absolute
magnitude, $M_X$,
in any particular band $X$, 
\(
M_{X} = -2.5 \log \int_{0}^{\infty} S_{X}(\lambda)\, L_{\lambda}\,
d\lambda + C_X
\),
where $S_{X}$ is the response of filter $X$, $L_{\lambda}$ is the
luminosity per unit wavelength, and $C_X$ is the zero point of filter
$X$ determined from standard stars. Then one immediately obtains a
distance modulus 
\( \mu_X \equiv m_X - M_X - A_X = 5\log{(d/10\textrm{pc})}, \)
where $m_X$ is the apparent magnitude in band $X$ and $A_X$ is the
extinction due to dust along the line of sight either in the host
galaxy or in our own galaxy.  The SEAM
method does not need to invoke a blackbody assumption or to calculate
dilution factors.

\section{Fits and Conclusions}

\begin{figure}[!ht]
\begin{center}
\includegraphics[height=4cm,angle=90]{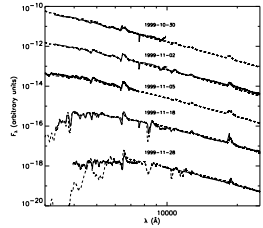}
\caption{\label{fig:sn99em_fits}The synthetic spectra are compared to
  observed spectra at 5 different epochs. The observed spectra were
  obtained at CTIO for Oct 30, Nov 2, and Nov 18 \citep{hamuyepm01}, at
  \emph{HST} and FLWO on Nov 5 \citep{bsn99em00} and the optical
  spectrum on Nov 28 was obtained at Lick \citep{leonard99em02} while
  the IR was obtained at CTIO \citep{hamuyepm01}. The observed fluxes
  have been offset for clarity.}
\end{center}
\end{figure}

Using the latest version of \phx\ version 13.06, we have calculated
synthetic spectra of SN~1999em using  Model S15 of 
\citet{ww95}. The 
model was expanded homologously and the gamma-ray deposition was
parameterized to be consistent with the nickel mixing found in
SN~1987A \citep{mitchetal87a01}. The abundances were taken directly from
the model, and the effects of radioactive decay were taken into
account. No attempt was made to adjust the metalicity or helium
mixing. The model had abundances: $X = 0.64$ $Y = 0.34$ $Z =
0.019$. Figure~\ref{fig:sn99em_fits} shows the results are quite close
to observed results except at the latest epoch where the blue is
strongly dependent on the exact abundances. With these models we
are able to obtain a reasonably accurate distance to SN~1999em
\citep{bsn99em04}. 

Results from both nearby and high-z searches require quantitative
spectroscopy to interpret their results.  Results from hydrodynamical
calculations also require quantitative spectroscopy to validate/falsify
their results. The SEAM method applied to both nearby and high-z
samples will yield interesting cosmological results.  SEAM provides a
route to obtaining cosmological parameters  completely \emph{independent} from
SNe Ia.

\acknowledgments We thank Doug Leonard and Mario Hamuy for helpful
discussions.  This work was
supported in part by by NASA grant NAG5-3505, NSF grants AST-0204771
and AST-0307323, an IBM SUR grant to OU and
NASA grants NAG 5-8425 and NAG 5-3619 to  UGA
PHH was supported in part by the P\^ole Scientifique de
Mod\'elisation Num\'erique at ENS-Lyon. This research used resources
of: the San Diego Supercomputer Center (SDSC), supported by the NSF;
the National Energy Research Scientific Computing Center (NERSC),
which is supported by the Office of Science of the U.S.  Department of
Energy under Contract No. DE-AC03-76SF00098; and the
H\"ochstleistungs Rechenzentrum Nord (HLRN).  We thank all these
institutions for a generous allocation of computer time.


\end{document}